\documentclass[conference]{IEEEtran}

\usepackage[utf8]{inputenc}
\usepackage[russian,english]{babel}

\usepackage{placeins}

\usepackage{graphicx}
\usepackage{amsfonts} 

\title{\LARGE \bf Exploring the Impact of Password Dataset Distribution on Guessing}

\author{Hazel Murray$^{1}$ and David Malone$^{2}$
\thanks{This work was supported by an Irish Research Council Government of Ireland Postgraduate Scholarship}
\thanks{$^{1}$H. Murray is an IRC Scholar with the Department of Mathematics and Statistics and the Hamilton Institute, Maynooth University, Ireland.
        {\tt\small hazel.murray@mu.ie}}%
\thanks{$^{2}$D. Malone is with the Department of Mathematics and Statistics and the Hamilton Institute, Maynooth University, Ireland. 
        {\tt\small david.malone@mu.ie}}%
}

\begin{document}
\IEEEoverridecommandlockouts 
\IEEEpubid{\makebox[\columnwidth]{978-1-5386-7493-2/18/\$31.00~\copyright{}2018
IEEE \hfill} \hspace{\columnsep}\makebox[\columnwidth]{ }}

\maketitle
\begin{abstract}
Leaks from password datasets are a regular occurrence. An organization may defend a leak with reassurances that just a small subset of passwords were taken. In this paper we show that the leak of a relatively small number of text-based passwords from an organizations' stored dataset can lead to a further large collection of users being compromised. Taking a sample of passwords from a given dataset of passwords we exploit the knowledge we gain of the distribution to guess other samples from the same dataset. We show theoretically and empirically that the distribution of passwords in the sample follows the same distribution as the passwords in the whole dataset. We propose a function that measures the ability of one distribution to estimate another. Leveraging this we show that a sample of passwords leaked from a given dataset, will compromise the remaining passwords in that dataset better than a sample leaked from another source.
\end{abstract}

\section{Introduction}
Passwords are integral to our online security. Yet a major complication in password security is the storage of large datasets of passwords: password leaks are regularly announced, and many occur that we never hear about. In addition to leaks, passwords can be comprised in other ways, for example via social engineering, phishing or keylogging. These latter examples are resource consuming for the attacker, as they usually require per-user effort. In this paper, we investigate whether a relatively small number of passwords exposed can jeopardize the security of the passwords remaining in that dataset. Our basis for considering this is that previous research shows that users often incorporate details into their passwords which reflect the nature of the site they are creating the password for \cite{malone2012investigating,inglesant2010true,shen2016user}. For example, users might include the website name in their passwords or include colloquial words relevant to the website domain. In addition, users choosing passwords will be subject to the same password composition policies (e.g. including symbols and numbers in passwords) and users of a particular service may have considerable demographics in common (e.g. language, geographical location) that lead to common password choices. This investigation will reveal the true extent to which a database of text based passwords is vulnerable when a subset of passwords is compromised.

In this work, we consider using the exact passwords used by the compromised users to attack other users, similar to the attack in \cite{malone2012investigating}.  Other research has invested interest in the attacker's ability to guess passwords by generating further passwords using techniques such as Markov chains \cite{durmuth2015omen}, deep learning \cite{hitaj2017passgan} and probabilistic context-free grammars \cite{weir2009password}. We focus on estimating the popular passwords in the distribution, and how knowing these improves returns on guessing effort. We will show that a small sample of leaked passwords provides us with enough insight into the overall distribution of users' passwords for tailored password guessing. 

In Section~\ref{sec:model} we first discuss the datasets we use for our empirical evidence. Then we describe the notation used in the paper, how we created the password distributions and our password guessing model. Section~\ref{sec:ethics} discusses the ethical considerations involved when using leaked password datasets. Section~\ref{sec:1} demonstrates the plausibility of using a leaked sample to guess other samples of passwords from the same password dataset. Section~\ref{sec:Sanov} shows the theoretical underpinnings supporting our ability to understand the distribution of the full dataset using a sample. Section~\ref{sec:ratioNtox} looks at the importance of the ratio of the number of passwords in the sample to the number of passwords in the dataset. Section~\ref{sec:5} explores the impact a leak of a subset of passwords from a dataset has on the security of the rest of the passwords. Section~\ref{sec:function} introduces our function which provides a metric comparing how one password distribution can estimate another. Finally, in Section~\ref{sec:applyfunction} we apply our guessing function to verify our claim that guessing a password dataset using a sample from that dataset is an efficient and effective method for compromising a large number of users' passwords. 


\section{Datasets, Guessing Model and Notation} \label{sec:model}

\subsection{Overview of datasets} \label{sec:datasets}
We collected  password datasets that had been compromised and were subsequently leaked to the public. The datasets  were compromised by various methods (e.g. key-logging, network sniffing or database dumps) so the lists may only contain a random, and possibly biased, sample of users.  The lists used in this paper are from rockyou.com, hotmail.com, flirtlife.de, and compubits.ie, and contain $32602877$, $7300$, $98912$ and $1795$ passwords respectively.
We cleaned up the datasets by taking a user's password as the last entry seen for that user and omitting any user with a whitespace password.  

\subsection{Notation} \label{sec:notation1}
We denote password datasets as $X$ and the number of passwords in this dataset $|X|$. We rank and order the passwords in the dataset to generate a distribution which we call $p_0$. We take $n$ passwords from 
this dataset and they make up one sample, denoted $q$. This sample is then ordered and used to guess other passwords. We refer to the act of using a sample to guess as a trial. If we complete multiple guessing trials $t$ using a sample $q$ we use a subscript of $q_t$. If we want to emphasize the size of the sample, $n$, we denote it with a superscript. For example $t$ trials of size $n$ would be denoted $q_t^n$.

\subsection{Model of distributions} \label{sec:notation2}
We choose $n$ users' passwords randomly with replacement from a password dataset and organize these by counting the number of times we see each password in the sample. For example, in the full rockyou dataset the password \textit{123456} is seen 290729 times \cite{malone2012investigating}. These rankings are then ordered from highest to lowest, thus becoming the distribution. We treat the first sampled distribution, $q_0$, as the users' passwords that are available in plaintext to an attacker. 

For example, a sample $q_0$ is plotted in Fig.~\ref{fig:1_N=M=100ThousandRockyou} for $n=100000$ passwords from the rockyou dataset and is visualized by plotting the number of users that would be compromised by guessing the $g$ most popular passwords (the upper pink line). The number of guesses, $g$, is naturally limited by the number of unique passwords that exist in the sample (in this case 82479). If we let $\sigma_{q_0}(g)$ be the password of rank $g$ in the sample $q_0$, then this distribution can be described as:
\begin{equation}
\mathcal{F}_{q_0}(g) = \sum_{k=1}^{g} {q_0}(\sigma_{q_0}(k))
\end{equation}
Guessing $q_0$ in this order gives us an upper bound on the number of user accounts compromised after guessing $g$ distinct passwords. 

\subsection{Model of Password Guessing}
We are interested in the ability of a sample of passwords leaked from a dataset to guess another sample of passwords from the same or another dataset. We construct a number of distributions, $q_t$, each with $n$ users' passwords. We measure the ability of $q_0$, the distribution we know, to guess the passwords in the other samples efficiently and effectively. 

With $q_0$ ordered from most prevalent to least, we guess the passwords in the new $q_t$ samples by taking the most popular password in $q_0$ and recording how many times it occurs in $q_t$, taking this as the number of successes\footnote{We are assuming an attacker can guess a particular password against many accounts easily. This could be via offline guessing of password hashes, online guessing not subject to rate limiting, etc.}. This is repeated for the top $g$ password guesses that we are willing to make. The function describing this guessing attack is:
\begin{equation}
\mathcal{G}_{q_t}(g) = \sum_{k=1}^{g} q_t(\sigma_{q_0}(k)).
\end{equation}

\section{Ethics} \label{sec:ethics}
As part of this study we collected password datasets that had been compromised and were subsequently leaked to the public. There arises an issue of privacy and security as a result of collecting and analyzing these password databases. We have used the current best practice to minimize any harm associated with using this data. This is an account of our ethical considerations in line with our Research Ethics Board and \cite{ethics}.
\paragraph{Stakeholders}
The stakeholders in this scenario are the users whose password has been included in the leaked password dataset. Also the organization from which the passwords were leaked. 
\paragraph{Informed consent}
It is not practical to gather consent from the stakeholders affected. The password datasets we use are already accessible to the public using common search engines. Our paper is not the first publication to reference these specific password datasets \cite{wang2016implications}, so we know the existence of the leaks is already known. 
\paragraph{Harms}
The passwords leaked could still be in use by individuals. The passwords themselves could contain personal information. In some cases the leaked database includes other personal details such as email addresses or names. 
\paragraph{Safeguards}
We removed the personally identifying information, usernames and emails, from our datasets before analysis. We recorded the frequency with which each password occurred in the database and then ranked these frequencies. This was the only information we needed to retain. Therefore the actual passwords leaked do not appear anywhere in our paper. (We mention that the password ``123456'' appeared 290729 times in the rockyou.com password dataset since this has been published previously by other researchers \cite{malone2012investigating} \cite{zhang2017password}.) 
\paragraph{Public interest}
Attackers have access to these password datasets and likely structure their attacks using the knowledge gathered from them. Therefore it is in the public interest for our analysis and defenses to be derived using the knowledge we can glean from these available password datasets. 
The use of these datasets by multiple researchers is positive for reproducibility and offers advantages over ``generated'' passwords created by participants in controlled studies \cite{fahl2013ecological}. 

\section{Guessing One Sample Using Another} \label{sec:1}
\begin{figure}[ht] 
	\includegraphics[width=\linewidth]{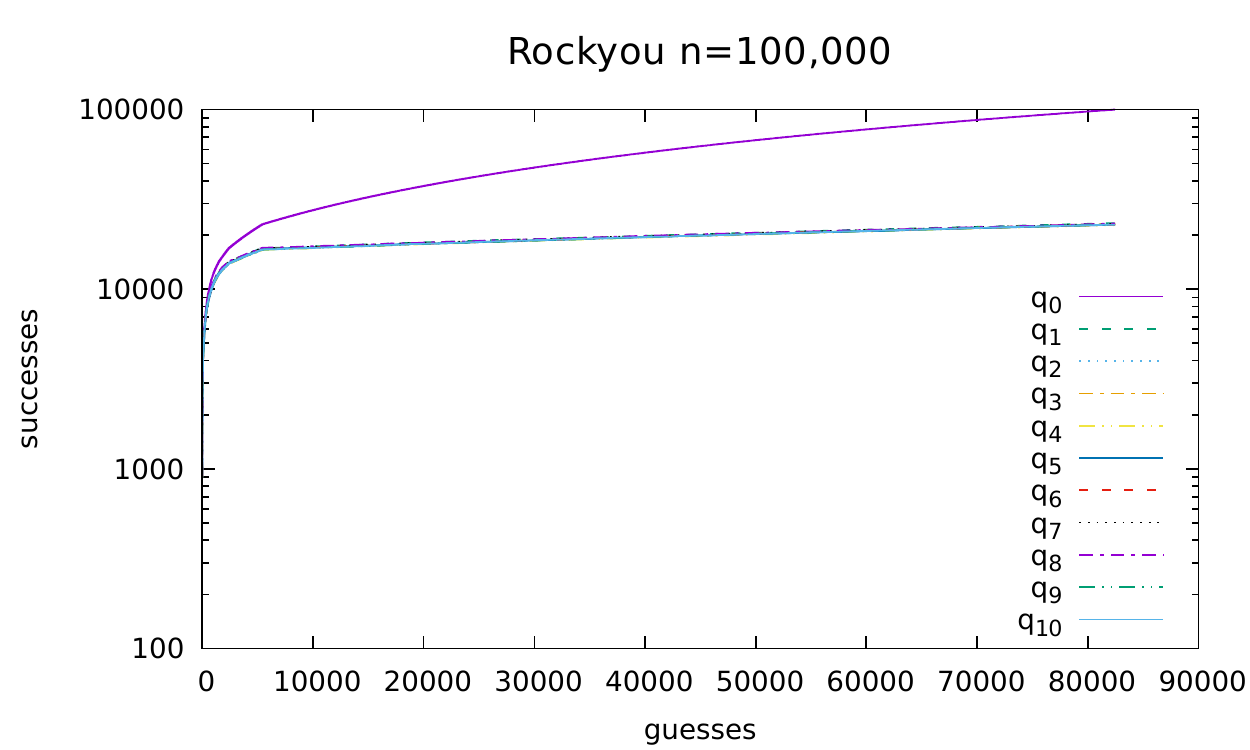}
\caption{Rockyou passwords, ten trials with sample size $n= 100000$, each guessed with the distribution of $q_0$.}
\label{fig:1_N=M=100ThousandRockyou}
\end{figure}

\begin{figure}[ht] 
	\includegraphics[width=\linewidth]{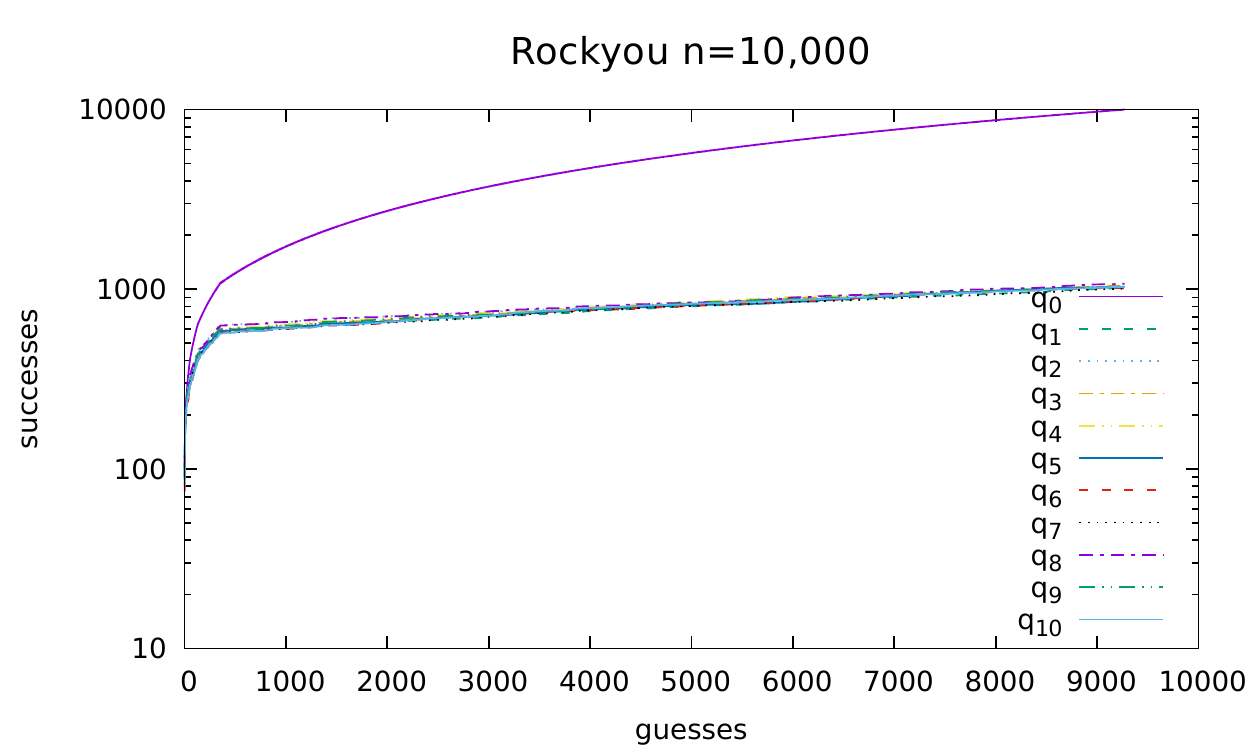}
\caption{Rockyou passwords, ten trials with sample size $n= 10000$, each guessed with the distribution of $q_0$.} \label{fig:2_N=M=10000Rockyou}
\end{figure}

\begin{figure}[ht] 
	\includegraphics[width=\linewidth]{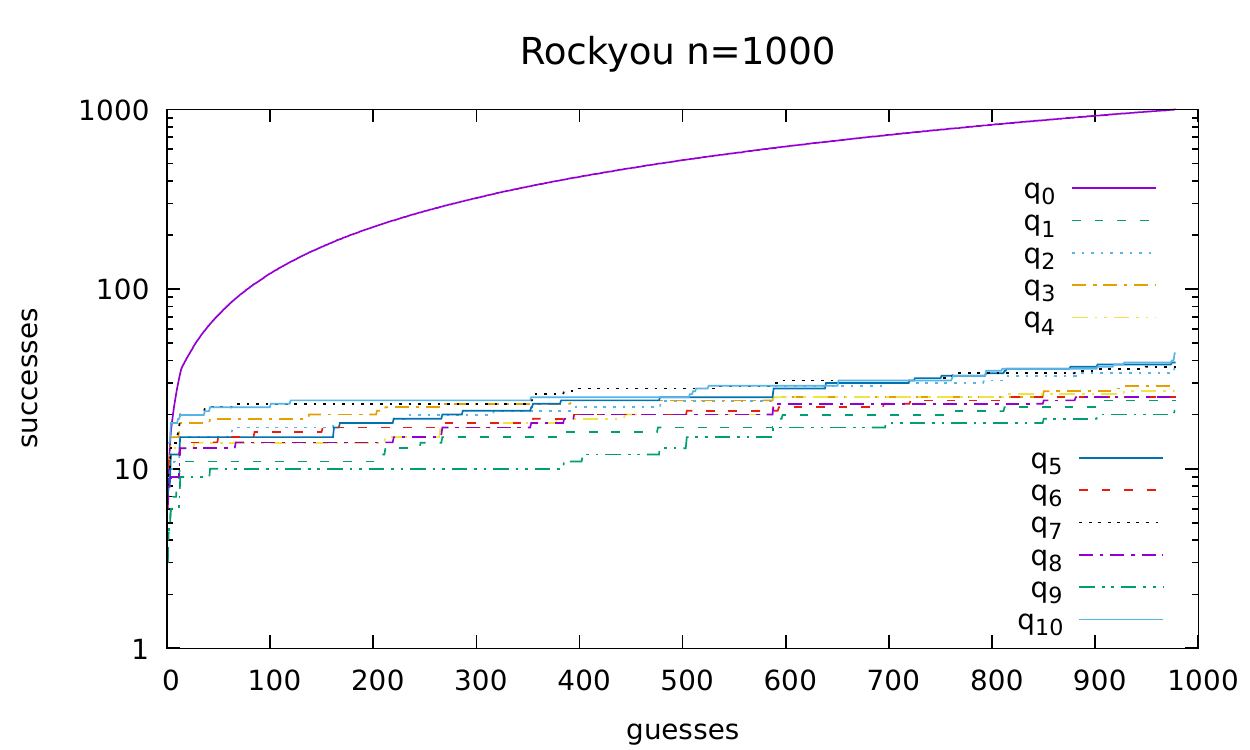}
\caption{Rockyou passwords, ten trials with sample size $n= 1000$, each guessed with the distribution of $q_0$.} \label{fig:3_N=M=1000Rockyou}
\end{figure}

\begin{figure}[ht] 
	\includegraphics[width=\linewidth]{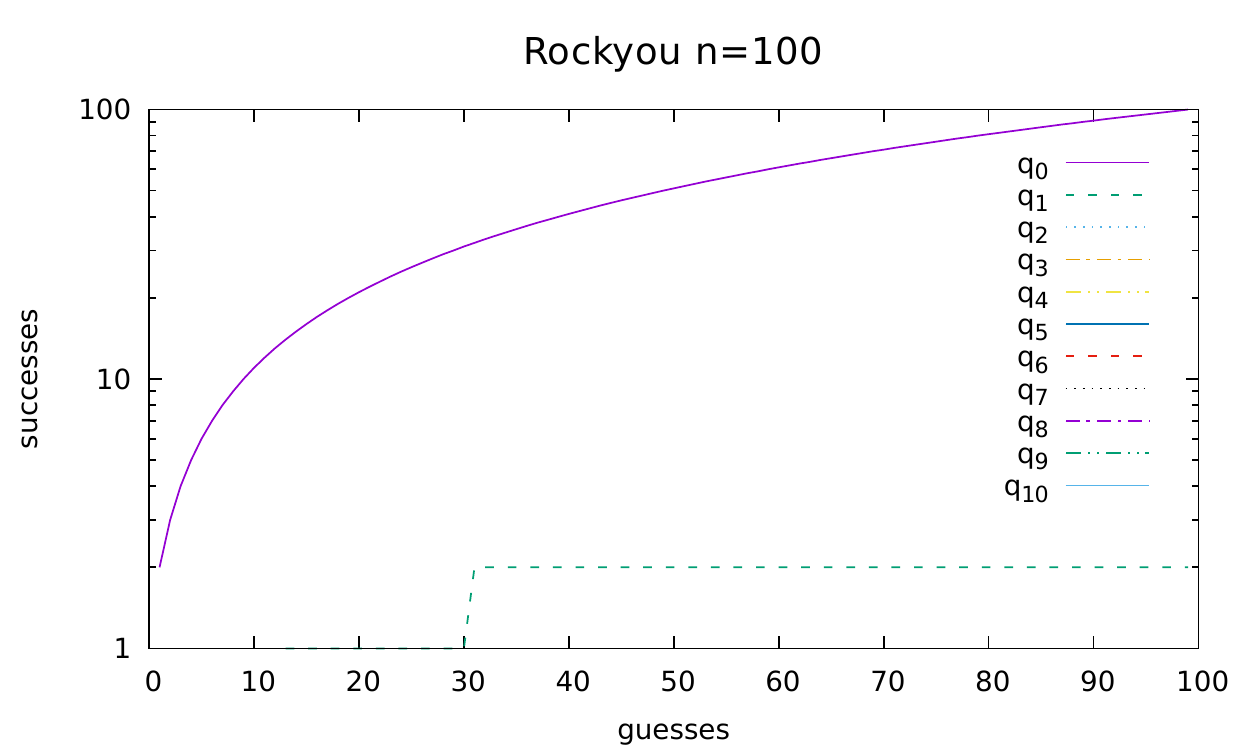} 
\caption{Rockyou passwords, ten trials with sample size $n = 100$, each guessed with the distribution of $q_0$. $q_0$: $\mathcal{F}_{q_0}(g)$ and $q_{t}$: $\mathcal{G}_{q_{t}}(g)$} \label{fig:4_N=MRockyou}
\end{figure}

First we  depict how well one sample of passwords taken from a dataset set can guess another sample taken from the same dataset. We look at this for four different sized samples; $n=$ 100, 1000, 10000 and 100000. For each number of samples we conduct ten trials to gauge how diverse the results are. Our plot for each sample size shows the ability of some sample, $q_0$, when ranked and ordered, to guess ten other samples, $q_1 \dots q_{10}$. We also include a line depicting how well $q_0$ guesses itself, $\mathcal{F}_{q_0}$, to provide a means of comparison. In Fig.~\ref{fig:1_N=M=100ThousandRockyou} -- \ref{fig:4_N=MRockyou} we take our samples from the rockyou dataset.

Fig.~\ref{fig:1_N=M=100ThousandRockyou} illustrates the ability of $q_0$ to guess $q_0 \dots q_{10}$ where each 
trial is describing a sample of size $n=100000$ taken from the rockyou dataset. We can immediately see that there is little variation between the distributions of $q_1 \dots q_{10}$. In fact, if we check $\mathcal{F}_{q_i}$ for $i = 1, \dots, 10$ they seem to follow a similar path to $q_0$. We see that at $g=10,000$ guesses our optimal number of successes, based on $q_0$, is approximately $27521$ and our number of successes for $q_t$ guessed with $q_0$ is approximately $17113$.
When all the passwords in $q_0$ were guessed ($g=82479$) against $q_t$ we had successfully guessed around $23020$ users' passwords (compared to 100,000 for $q_0$). We can also see that we were able to compromise a large number of users with relatively few guesses; notice the jump at the beginning of the graph. 

Now consider trials with a smaller number of passwords, $n= 10000$. Fig.~\ref{fig:2_N=M=10000Rockyou} shows a broadly similar result to Fig.~\ref{fig:1_N=M=100ThousandRockyou},  however the $q_t$ distributions overlap less. At $g=1000$ guesses we have compromised 600 users' passwords in $q_6$, the lowest lying distribution. The optimal at $g=1000$ is $\mathcal{F}(g)=1727$.

Fig.~\ref{fig:3_N=M=1000Rockyou} shows the results with $n = 1000$. It shows an increased amount of variation between the $q_t$ distributions. However they do still follow a similar shape to $q_0$. We also still manage to achieve most of our successes in the first few, $g < 20$, guesses.  

Fig.~\ref{fig:4_N=MRockyou} shows that when the number of passwords in the sample is too small ($n = 100$) we cannot glean enough information from one sample to effectively guess another sample of a similarly small size. In fact, we conducted supplementary experiments by taking samples with $n=100$ passwords for $q_0$ and using an increasing number of passwords for the $q_t$ trials that we were guessing. These results showed that we were unlikely to guess more than 100 users' passwords, even when the sample we were guessing had $n=10000$ users' passwords. We believe this result reflects the existence of a \emph{heavy tail} of low-frequency passwords in many password datasets \cite{malone2012investigating}. Even the rockyou dataset has a large number of passwords with frequency 1.

These results show that the effectiveness of guessing passwords from a dataset is dependent on the number of passwords, $n$, in the sample. It seems likely that the guessing ability of $q_0$ is not only affected by $n$ but is actually dependent on the relative size of $n$ to the size of the total dataset, $|X|$. We will investigate this further in the next two sections.

\section{Sanov's theorem} \label{sec:Sanov}
We are interested in the likelihood that the distribution of the sample chosen reflects the distribution of the passwords in the whole password dataset, so that the ranking of passwords for guessing is close to correct. Sanov's theorem \cite{sanov} \cite{russianSanov} gives a bound on the probability of observing an atypical sequence of samples from a given probability distribution, $p_0$. The definition of a typical set follows from the Asymptotic Equipartition Property \cite{cover2012elements}. A typical set $A^{n}_{\epsilon}$ with respect to $p_0$ is the set of sequences $(x_1,x_2,\dots, x_n) \in p_{0}^{n}$  with the property that $2^{-n(H(x)+\epsilon)} \leq p(x_1,x_2,\dots, x_n) \leq 2^{-n(H(x)-\epsilon)}$ for $H$ the information entropy of $x$ and $\epsilon > 0$. The probability that the empirical distribution $q^n$ of size $n$ falls within a given set $A$ is bounded by
\begin{equation}
\mathbb{P}[q^n \in A] \leq (n+1)^{|X|} 2^{-nD_{KL}(p^*||p_0)}
\end{equation}
where $p^* \in A$ is the distribution that gives the smallest distance between $p_0$ and $A$ measured using the Kullback-Leibler divergence, $D_{KL}$. The Kullback-Leibler divergence is a measure of the information lost when one probability distribution estimates another \cite{kullbackkullback}. For our case we take $A$ to be probability distributions $q^n$ where the Kullback-Leibler divergence is greater than a constant $\alpha > 0$. We consider these distributions to be atypical. This leaves us with the result that
\begin{equation}
\mathbb{P}[q^n \in A] \leq (n+1)^{|X|} 2^{-n(\alpha)} \longrightarrow 0
\end{equation}

So the probability that we see an atypical distribution tends towards zero. Implying that the probability that the distribution of our samples is typical (i.e. is a small distance from the distribution of the real dataset) tends towards 1 as the sample size $n$ becomes large. This shows that for sufficiently large $n$ with respect to $X$, our samples drawn will follow a similar distribution to that of the overall distribution they were drawn from. 

Finding the turning point of this probability function tells us the point at which the function starts decreasing towards zero. We take the log of the function and then differentiate to get:
\begin{equation}
\frac{|X|}{n+1} - \alpha \ln 2 = 0
\end{equation}
This confirms that the ratio of the sample size to the total size of the dataset is relevant to the prevalence of atypical samples when we take passwords from our datasets. We explore this further empirically in the next section. 

\section{Exploring the Relative Sample Size} \label{sec:ratioNtox}
In this section we use real password datasets to investigate the ratio between the number of passwords in the sample, $n$, and the total number of users' passwords in the dataset, $|X|$, and this ratio's impact on the ability of $q_0$ to estimate $q_t$ for $t=1,2,\dots,10$.

We will explore this using a constant $n = 1000$ users' passwords and observe the ability of $q_0$	to guess $q_t$  when the sizes of the full datasets were $|X|=1795$, $7300$, $98912$ and $32602877$ respectively for compubits, hotmail, flirtlife and rockyou.

\begin{figure}[ht] \hspace{-1em}
\begin{tabular}{c c c}
\includegraphics[width=0.49\linewidth]{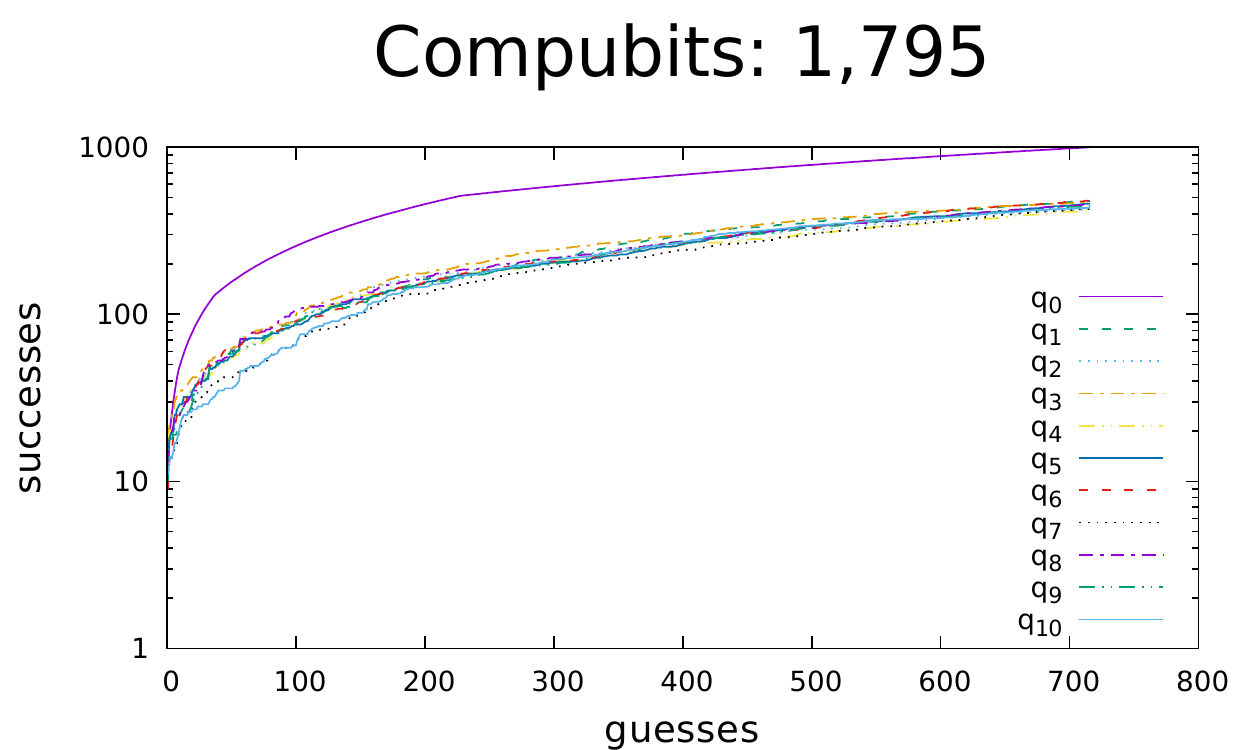} & \includegraphics[width=0.49\linewidth]{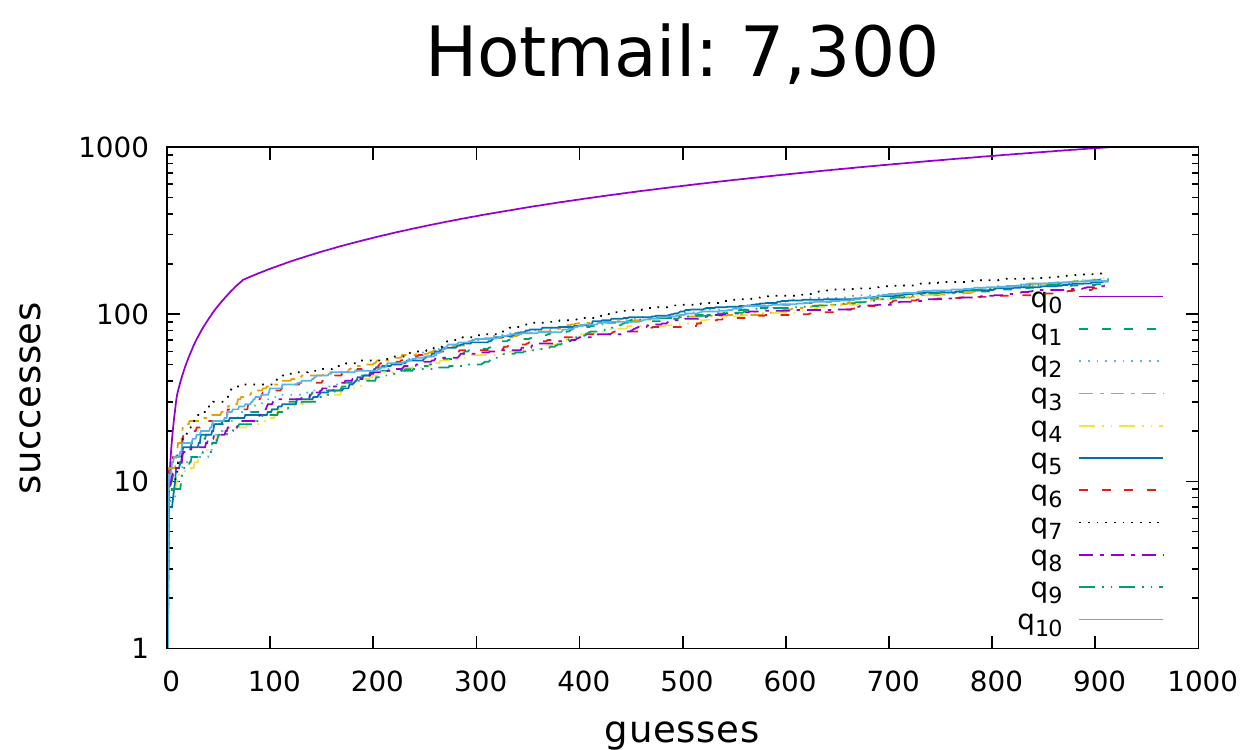} \\
	\includegraphics[width=0.49\linewidth]{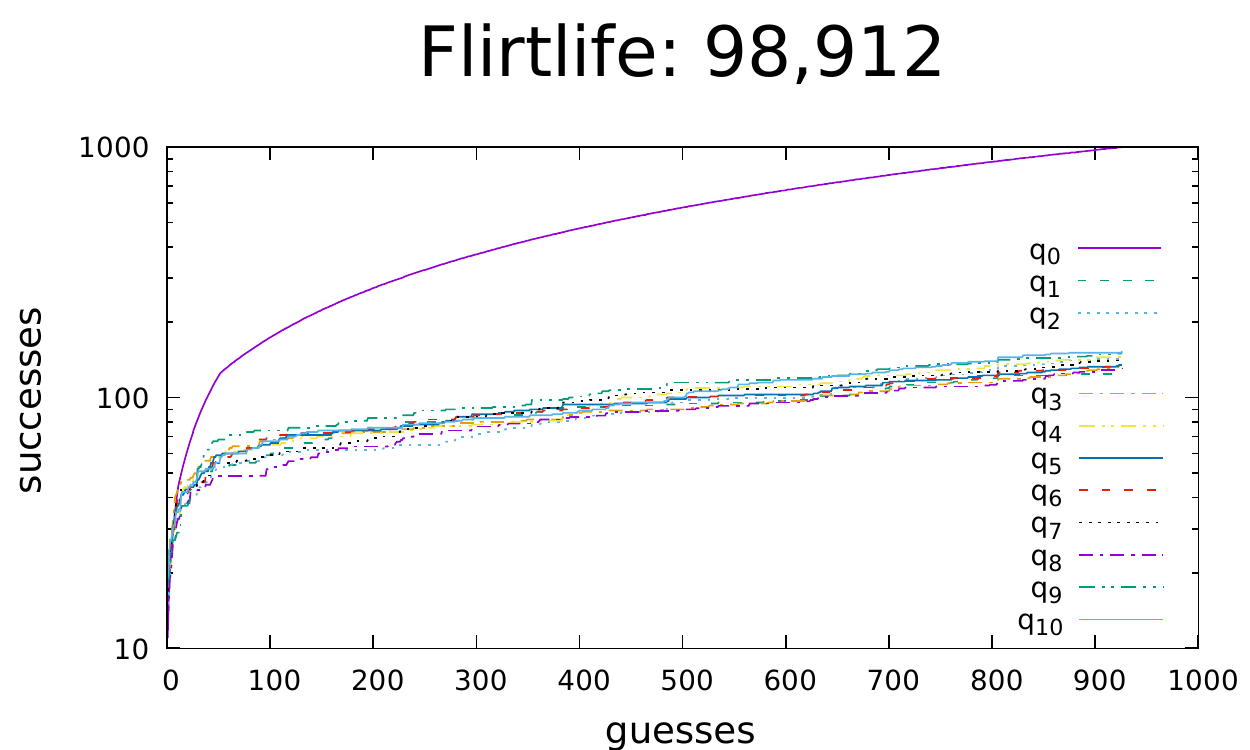} & \includegraphics[width=0.49\linewidth]{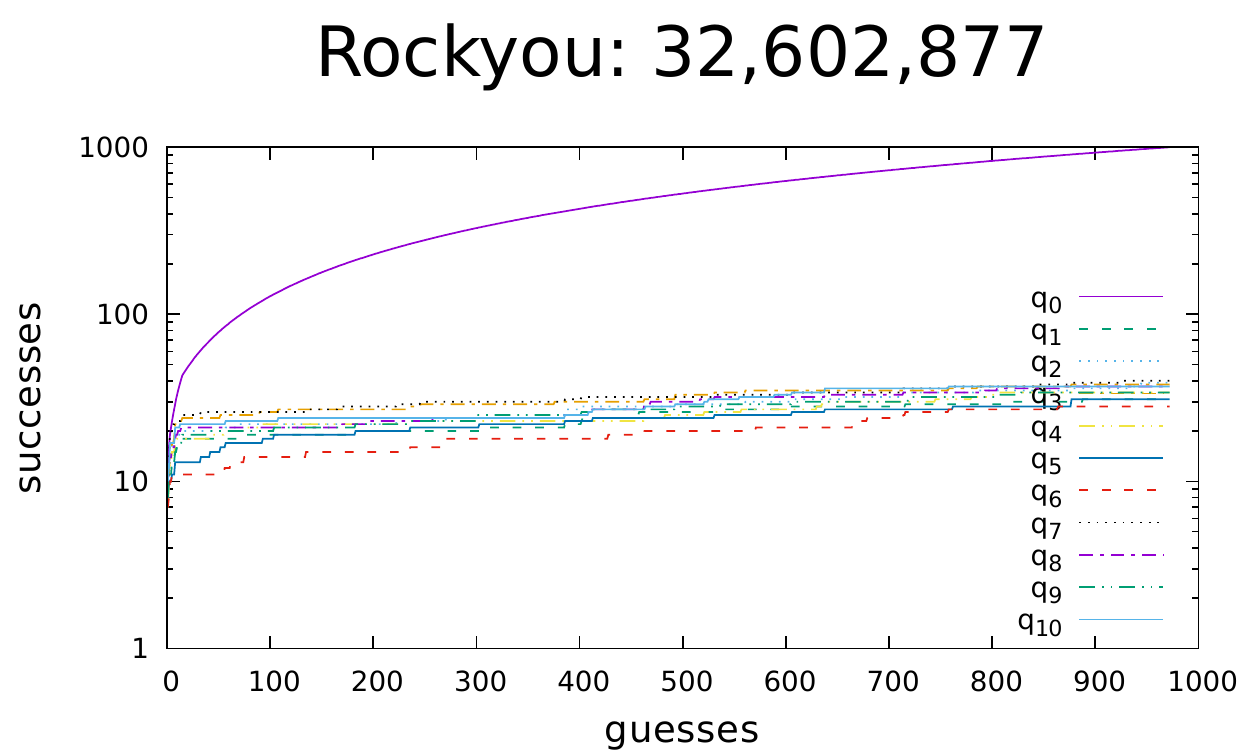} 
\end{tabular} \caption{Samples of size $n=1000$ taken from Compubits, Hotmail, Flirtlife and Rockyou.}
\label{fig:|X|_N=1000}
\end{figure}

Fig.~\ref{fig:|X|_N=1000} plots the ability of a $n=1,000$ sample to guess ten other $n=1000$ samples drawn from the same dataset. The samples in each case are ranked and ordered from most frequent to least frequent. We repeat this in Fig.~\ref{fig:|X|_N=1000} for our four password sets. 

In the graph of Compubits, $n=1000$ and $|X|=1795$, perhaps unsurprisingly we see the gap between optimal guessing (top line) and the guessing rate for each sample is the smallest of the four experiments. Taking a relatively modest number of guesses, $g=100$, we find that the lowest value for $\mathcal{G}(100)$ from the ten trials was 70, 24, 53 and 14 for Compubits, Hotmail, Flirtlife and Rockyou respectively. We can see that a sample of $n=1000$ from the smallest dataset, $|X|=1795$, compubits allowed for better guessing than a similar sample from the largest dataset, $|X|=32602877$, rockyou.

This provides empirical evidence that the smaller the difference between the number of passwords in each sample, $n$, and the total number of passwords, $|X|$, the more accurately we can guess.  

\section{Using a sample of size $n$ to guess the rest of the password dataset}  \label{sec:5}
In the scenario that an attacker gains access to a relatively small number of passwords through social engineering or some other attack vector, the goal of the attacker could be to compromise a large number of accounts from the same dataset using the passwords they have. 

\begin{figure}[ht]
      \centering
      \includegraphics[width=\linewidth]{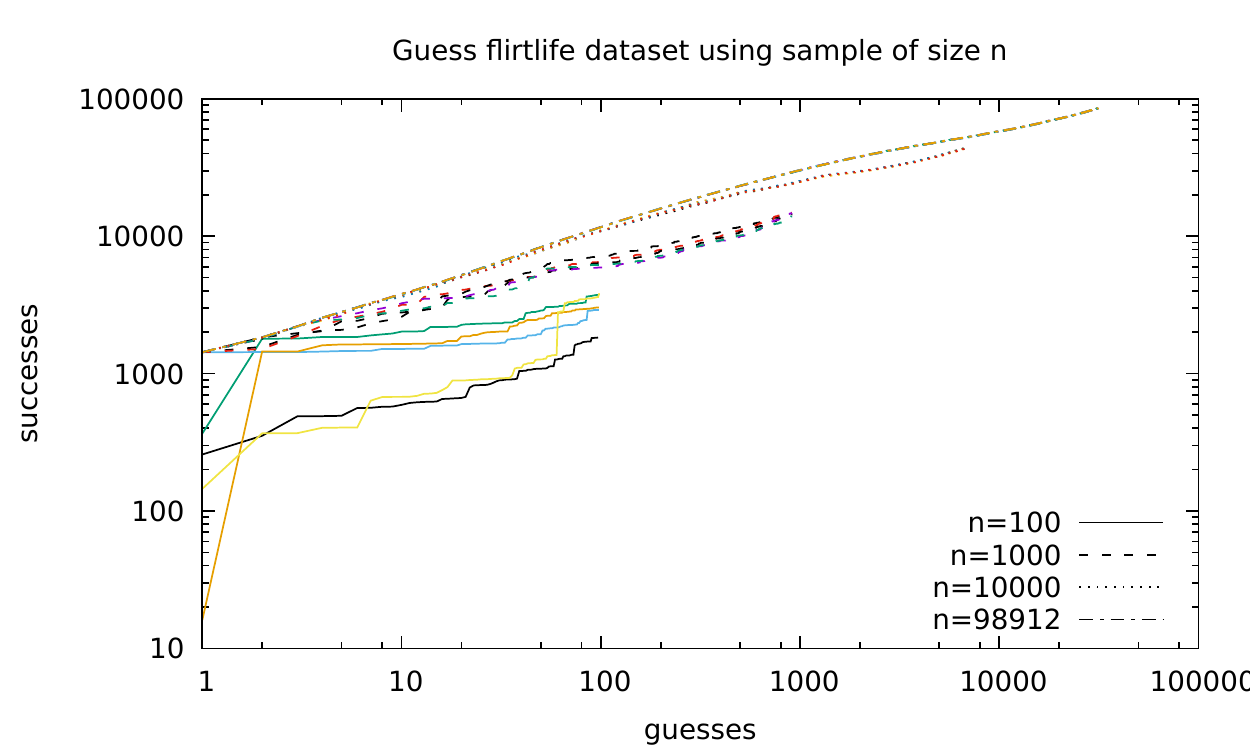}      
      \caption{Uses $n=100$, $n=1000$, $n=10000$ and $n=98912$ sized samples to guess the $|X|=98912$ passwords in the Flirtlife dataset.}
      	\label{fig:guessX}
   \end{figure}

We look at whether an attacker with $n$ users' passwords, chosen from the flirtlife dataset randomly with replacement, can use these to compromise a further larger number of users' passwords. 
Fig.~\ref{fig:guessX} takes samples of size $n=100$, $n=1000$, $n=10000$ and $n=98912$ from the Flirtlife dataset. Ranking and ordering these password samples to form distributions, we try to guess the whole password dataset which contains $|X| = 98912$ passwords. We repeated this five times for each of the different $n$ sample sizes to support an awareness of the spread the data can yield.

When $n=100$ users' passwords this represents taking $n/|X| \rightarrow 0.1\%$ of the Flirtlife dataset. Guessing every unique password in this sample against all the passwords in Flirtlife resulted in between $1829$ and $3800$ successful guesses for our five trials.  These represent between 1.8\% and 3.84\% of passwords in $X$ successfully compromised.

A sample including $n=1000$ users' passwords was $n/|X| \rightarrow 1.01\%$ of the Flirtlife dataset. Using this sample, the lowest one of the five trials yielded $13955$ and the highest yield was $14835$.
That is, between 14.1\%  and 15\% of users compromised. 

The sample of size $n=10000$ users' passwords represented $n/|X| \rightarrow 10.1\%$ of the Flirtlife dataset. This sample was able to guess between $43787$ and $44247$ users' passwords, which is 44.26\% to 44.7\% of the Flirtlife users respectively.

The final sample of size $n=98912$ represents $n/|X| \rightarrow 100\%$ of the Flirtlife dataset. However, because we choose samples from the dataset randomly with replacement, none of our five trials managed to successfully guess every password. We will revisit this idea in Section \ref{sec:applyfunction}. The lowest of the five trials yielded $85107$ and the highest yielded $85421$, representing between 86\% and 86.3\% of the passwords in the dataset. It is interesting to note that the highest number of unique passwords gathered by any of these five trials was $31427$, of a possible $43936$ unique passwords in the dataset. This plot offers a nice method for visual comparison of the different sample sizes. It emphasizes how closely the $n=10000$ trials were following the distribution of the $n=98912$ trials; in particular up until $g=80$ guesses.


Noting the log scale on both x- and y-axes of Fig.~\ref{fig:guessX}, we can see that the first guesses had high returns for successes. For $g=10$ guesses the attacker compromised a minimum of $\mathcal{G}(10) = 592$, $2591$, $3603$ and $3754$ users' passwords for $n=100$, $1000$, $10000$ and $98912$ respectively. Thus, the attacker can use the content of a sample to guess the whole dataset with considerable efficiency. 

This plot shows us the number of users we managed to compromise. Similar to how it was useful to compare the smaller sample sizes to the $n=98912$ sample, it could be useful to analyze the effectiveness of guessing by comparing how many users' passwords are compromised relative to the optimal guessing of those passwords. 

\section{Guessing function} \label{sec:function}
We want a metric that can quantify this ability of the sample to guess $p_0$, the real distribution of the whole dataset.
Using the two functions $\mathcal{F}_{q^n}(g)$ and $\mathcal{G}_{q^n}(g)$ we have already defined, we propose a function, $\mathcal{H}_{q^n}(g)$, to measure the effectiveness of guessing one distribution given knowledge of another sample distribution,
\begin{equation}
\label{guessing}
\mathcal{H}_{q^n}(g) = \sum_{k=1}^{g} p_0(\sigma_{p_0}(k)) - p_0(\sigma_{q^n}(k)).
\end{equation}
This function measures the ability of the sample $q^n$ to guess $p_0$ in $g$ guesses relative to guessing in the optimal order. Below, we will illustrate this guessing function by demonstrating two examples using the hotmail dataset. 

 \subsection{Guessing function using a sample of 7300 users' passwords}
\begin{figure}[ht]
      \centering
      \includegraphics[width=\linewidth]{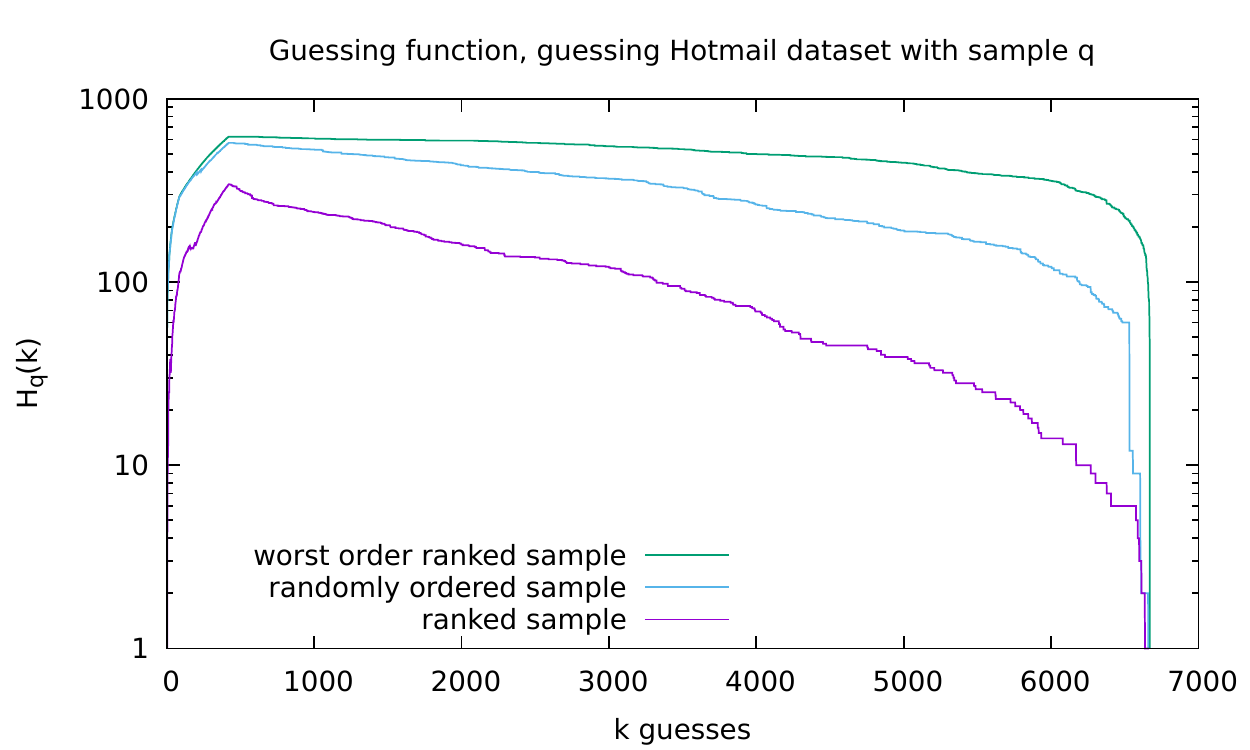} 
      \caption{Hotmail dataset guessed using a sample $q$, in best order, then in the worst order, then in a random order. $\mathcal{H}_{q_t}^{7300}(g)$}
      	\label{fig:Hotmail_function}
   \end{figure}

Fig.~\ref{fig:Hotmail_function} plots our guessing function for the effectiveness of a sample of $n=|X|=7300$ passwords chosen with replacement, to guess $X$ the hotmail dataset. Note, when graphing the function the lower values represent better guessing, since we want the minimum difference between the optimal guessing function, $\mathcal{F}_{q^n}(g)$, and our guessing with the sample, $\mathcal{G}_{q^n}(g)$. In Fig.~\ref{fig:Hotmail_function} we construct our trials in three different ways to demonstrate the function; best order, worst order and randomly ordered.

Because the sample size is the same as the size of the password dataset, we should guess nearly all the passwords in the dataset. This is portrayed by our function decreasing to zero for all three trials. 

First, we randomly choose $n=7300$ passwords from the Hotmail dataset and rank them in decreasing order of frequency, as we have done for all our trials up until this point. We can see that the distance our guessing is from the optimum increased rapidly in the beginning before stabilizing between $g=420$ and $g=422$, with an output of $\mathcal{H}(420) = \mathcal{H}(422)=342$, and then decreased down to zero.

Next, we take the distribution of the $n=7300$ samples but use the worst possible order. We use the same rankings of the passwords as in the previous trial but this time we reverse the order, guessing from least probable to most. This does indeed prove to be the least efficient method out of the three with most passwords compromised with the last few guesses. The turning point occurs at $\mathcal{H}(420) = \mathcal{H}(480) =622$. 

Finally, we take our $n$ samples and guess them in a random order. This random order has a turning point between $\mathcal{H}(420) = \mathcal{H}(455) = 574$. This was between the worst and best order for returns on guessing. 

We notice that each of our three distributions stabilize at $g=420$ before beginning to decrease towards zero. Looking at the password set we notice that only 420 passwords have a frequency higher than 1. So the remaining 6250 unique passwords all have frequency 1. This means that after $g=420$ the ``best order" trial is guessing as slowly as possible so the random and worst order sets gain during this point to all end at zero. We could expect our graph of the worst order to therefore plateau at this point, i.e. show no change when $420 \leq g \leq 6250$. However this is not the case as we have randomly chosen a sample of 7300 from the hotmail dataset set and therefore passwords can occur in the sample with frequencies different to those in the real password set. 

\subsection{Guessing function using 7300 users' passwords sampled without replacement}
\begin{figure}[ht]
      \centering
      \includegraphics[width=\linewidth]{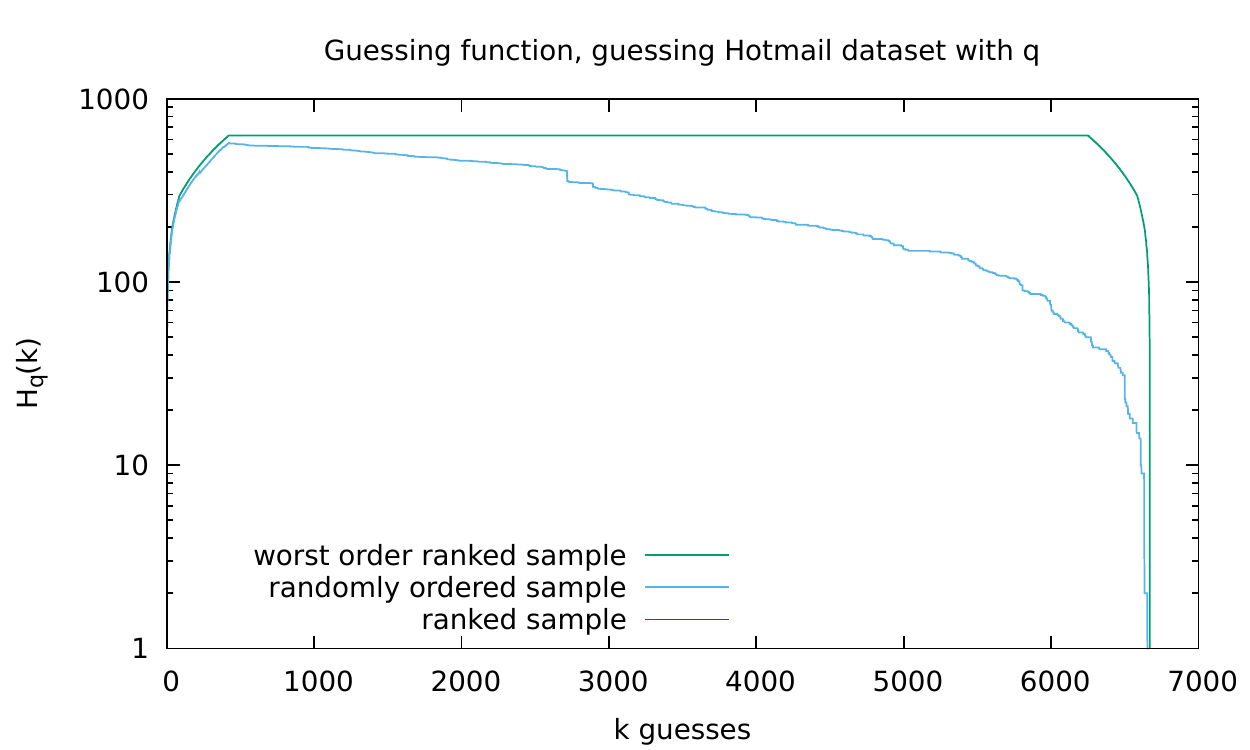} 
      \caption{Hotmail dataset guessed using all 7300 passwords taken from the hotmail dataset. Guesses in the best order, the worst order, then in a random order. $\mathcal{H}_{q_t}^{7300}(g)$}
      	\label{fig:Hotmail_function_wo_replace}
   \end{figure}

In Fig.~\ref{fig:Hotmail_function_wo_replace} we show the output of our guessing function for the best, worst and randomly ordered distributions created from the 7300 passwords in the hotmail dataset when they are chosen without replacement. 
When we rank and place the passwords in order from most popular to least, our guessing function $\mathcal{H}(g)$ returns zero at every guess $g$ since this is the optimal guessing order for guessing each password in the dataset. 

When we rank and order the passwords from the least popular to the most popular and guess them in that order we can see clearly where the tail of the distribution begins. At $g=420$ the distribution stabilizes and remains stable until we begin to encounter the passwords at the end of this distribution with frequency greater than 1. Specifically, we encounter this at: $\#unique.passwords - 420 = \#passwords.with.frequency.1 = 6250$. After this point the high frequency passwords cause the distribution to rapidly decrease to zero.

Our guessing function allows us to identify characteristics within our guessing methods and compare between different guessing strategies even when sample sizes differ.

\section{Effectiveness of Samples at guessing} \label{sec:applyfunction}
One of our initial claims was that revealing a sample of passwords from a dataset could help an attacker more than generic information about password use. In this section we provide evidence for this claim by using samples from one dataset to guess passwords in a different dataset. 

 \subsection{Guessing password datasets using samples from other datasets.}
 

We took samples of size $n=1000$ users' passwords from each of our four datasets. We then used these passwords to guess each of the datasets in full. We would have preferred to use a larger sample size, for example 10000, however we were limited by the number of passwords in the compubits password set.

\begin{figure}[ht]
      \centering
      \includegraphics[width=\linewidth]{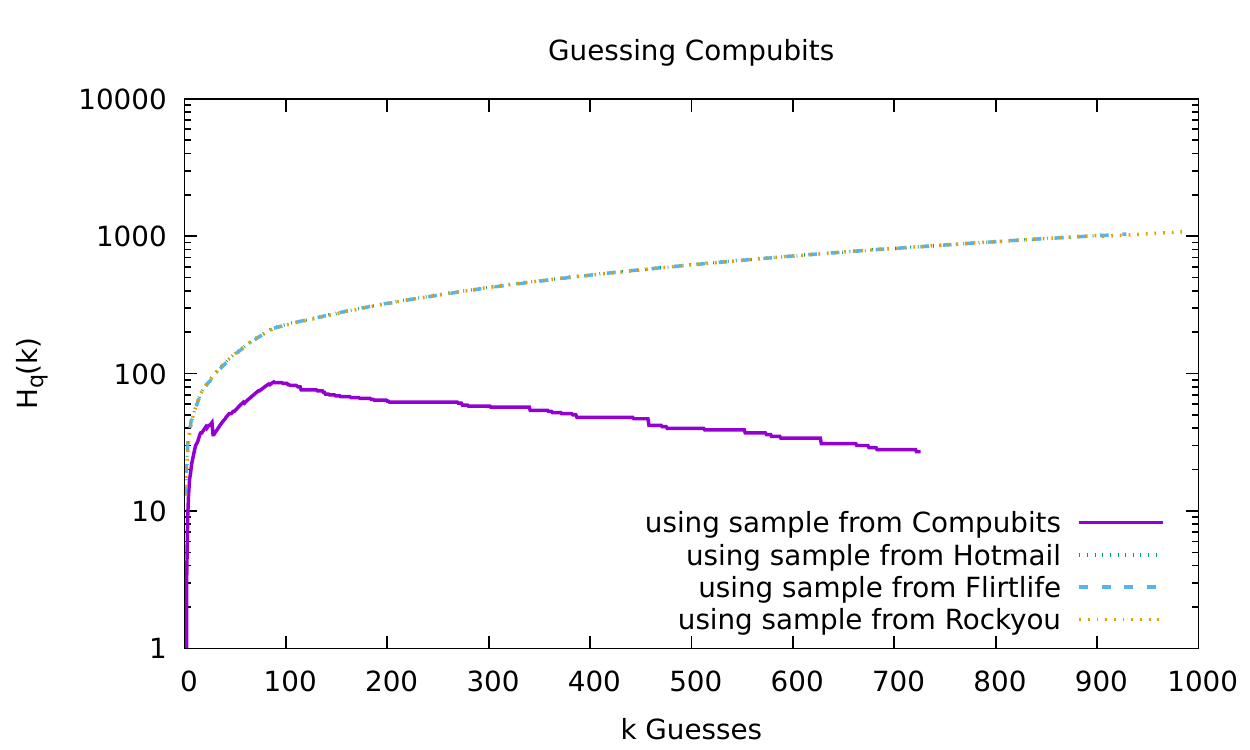}    
      \caption{Compubits dataset set guessed with $n=1000$ samples from compubits, hotmail, flirtlife and rockyou.}
      	\label{fig:compu_comp}
  \end{figure}

Fig.~\ref{fig:compu_comp} demonstrates that the compubits sample is the most effective sample for guessing the compubits dataset. In the sample of 1000 passwords taken from the compubits dataset, only 726 of them were unique. This is in comparison to 906, 929 and 984 from the hotmail, flirtlife and rockyou datasets respectively. This is interesting as it seems to not directly relate to the proportion that are unique in the whole dataset. \cite{malone2012investigating} reports that the $\frac{\#users}{\#unique.passwords}$ for each of the datasets is 0.92 for compubits, 0.91 for hotmail, 0.44 for flirtlife and 0.44 for rockyou. 
Looking at Fig.~\ref{fig:compu_comp} we can see that the graph showing compubits guessed using a sample from compubits,  begins decreasing after $g=88$. This is in line with our discussion in \ref{sec:function} which describes a turning point in our guessing when we encounter the ``tail of the distribution". In the case of the compubits dataset, we notice that after $g=88$ all passwords guessed have rank 1. We notice $\mathcal{H}(g)$ decreasing in the compubits graph because the sample of $n=1000$ passwords makes up a significant portion of the $|X|=1795$ password dataset. Therefore we reach this tail of the distribution within our 726 guesses and start our decline towards zero. 

\begin{figure}[ht]
      \centering
      \includegraphics[width=\linewidth]{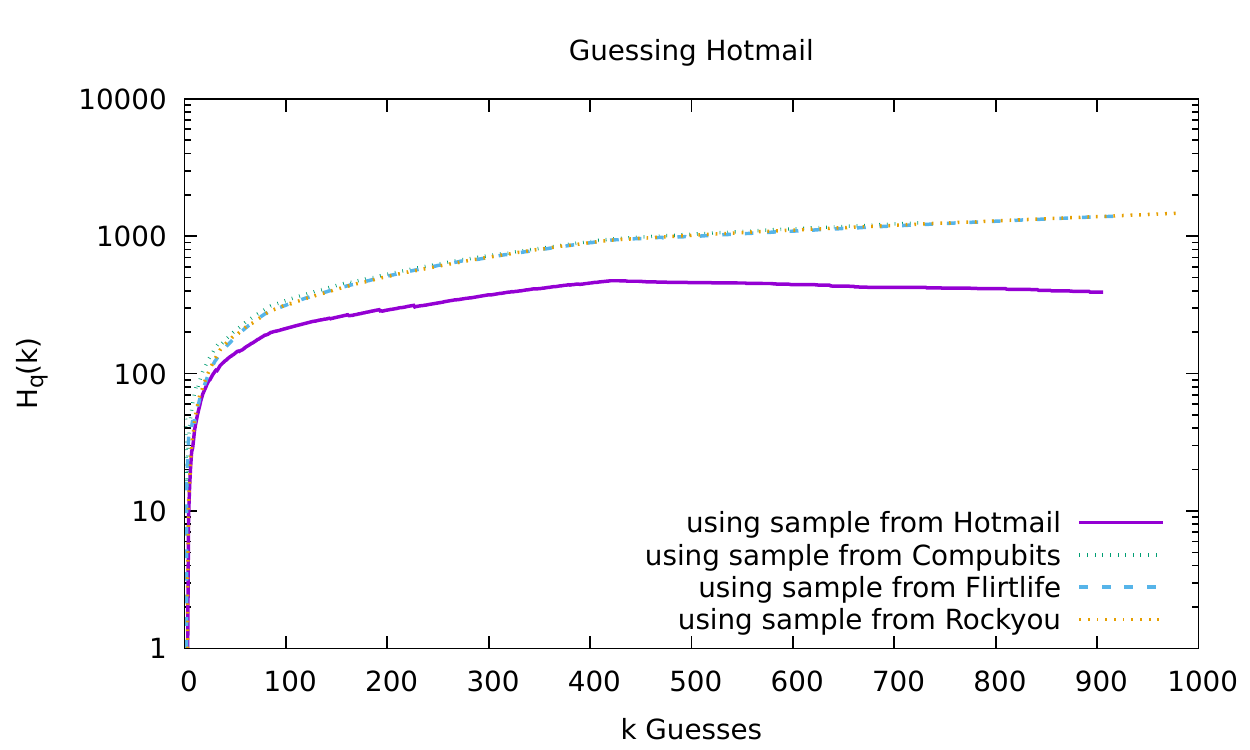}      
      \caption{Hotmail dataset set guessed with $n=1000$ samples from compubits, hotmail, flirtlife and rockyou.}
      	\label{fig:hot_comp}
  \end{figure}

Fig.~\ref{fig:hot_comp} shows that the hotmail sample guesses the hotmail dataset most effectively. We can also see that the distance from the optimum begins decreasing from $g=420$ guesses, where we encounter the passwords that occur with frequency 1. 

\begin{figure}[ht]
      \centering
      \includegraphics[width=\linewidth]{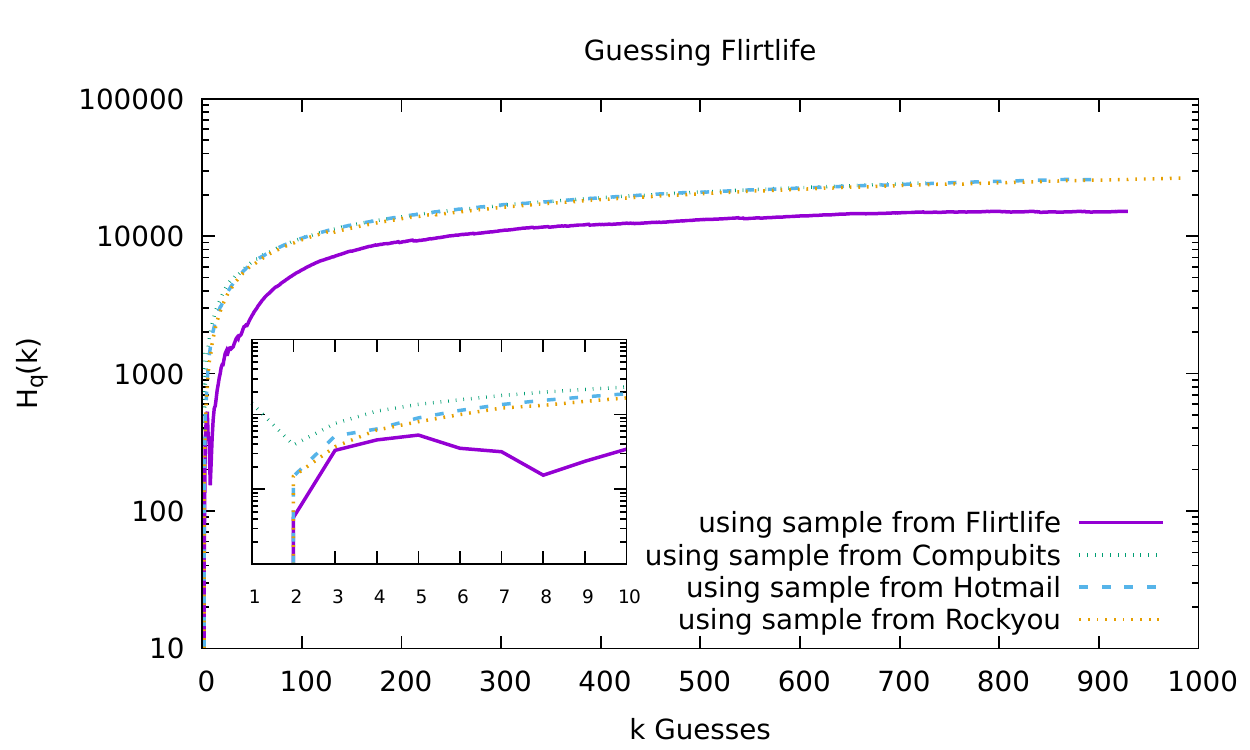}   
      \caption{Flirtlife dataset set guessed with $n=1000$ samples from compubits, hotmail, flirtlife and rockyou. With x range 1 to 10 expanded.}  
      	\label{fig:flirtlife_comp}
  \end{figure}

Fig.~\ref{fig:flirtlife_comp} depicts the flirtlife sample guessing the flirtlife dataset the most effectively. After approximately $g=150$ guesses we can see that little change occurs in the difference between the guessing ability of the different samples. This implies that any differences between the guessing ability of the samples are captured within the first few ``high frequency" password guesses. In this figure we have expanded the 1 to 10 x-range to show the rewards for these first few high frequency guesses. We can see that $g=8$ results in a high number of passwords guessed and seems to act as the defining factor in the difference between guessing using the flirtlife sample and the samples from the other datasets. 

\begin{figure}[ht]
      \centering
      \includegraphics[width=\linewidth]{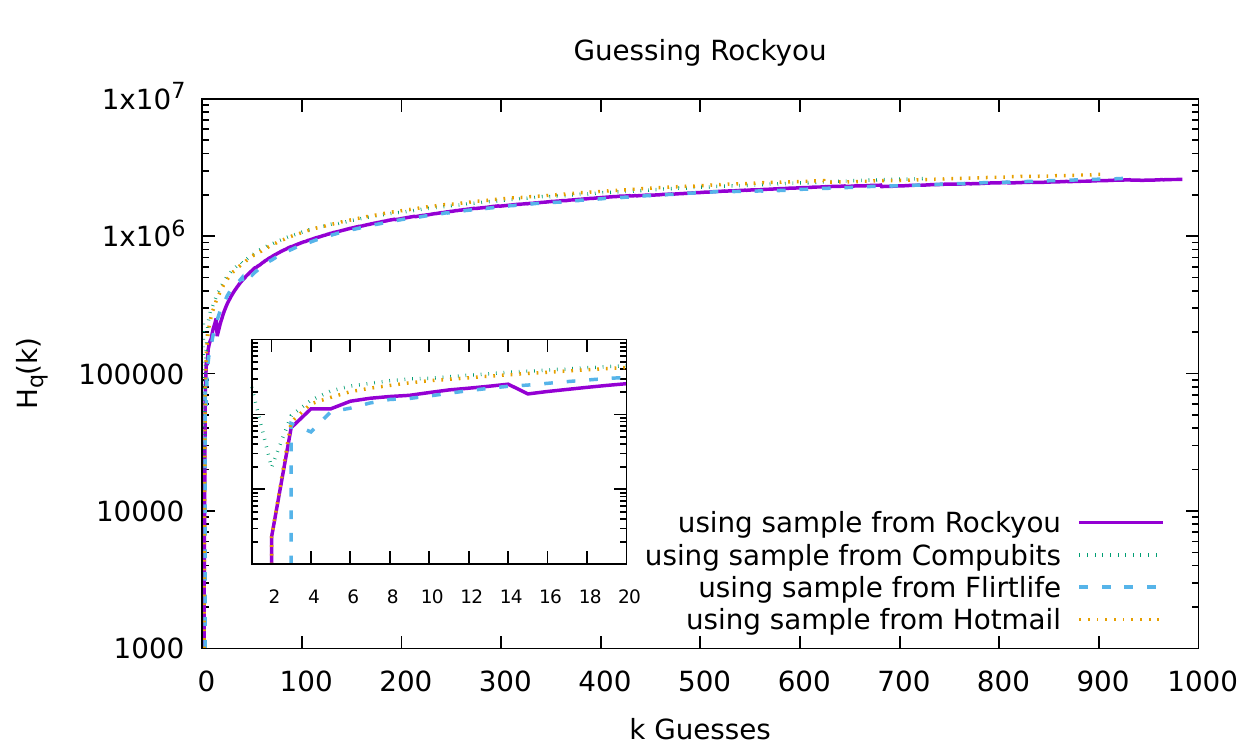}   
      \caption{Rockyou dataset set guessed with $n=1000$ samples from compubits, hotmail, flirtlife and rockyou.  With x range 1 to 20 expanded.}
      	\label{fig:rockyou_comp}
  \end{figure}

Fig.~\ref{fig:rockyou_comp} shows the rockyou password set guessed using each of our four samples. We notice that the compubits and hotmail samples are least effective at guessing the rockyou dataset. The large scale, $1$ to $1\times10^7$, on the y-axis hides the extent to which these samples are less effective but by the end of the guessing each of the two samples had returned over 200000 less passwords than the rockyou sample. 

There is very little difference between the returns from the fliftlife sample and the rockyou sample. Zooming in on the 1 to 20 x-range we see that the rockyou sample achieves better returns that the flirtlife sample at $g=15$ but they switch at $g=46$ when flirtlife is more effective. In fact throughout the guessing they switch subtly to become slightly more or less effective. After each had guessed each of the unique passwords in their samples, the rockyou sample had a distance from the optimal of 2596565 passwords and the flirtlife sample had a distance of 2633326; making the rockyou sample more effective at guessing by 36761 passwords. 

 Overall, we have observed that the sample taken from the dataset could guess it the most effectively. 
 
 \subsection{Guessing passwords in a set using samples chosen without replacement.}
It is possible that this result is a reflection of the fact that by taking $n$ samples from the dataset, we know there are at least those $n$ users' passwords in that dataset. 
To test if this explains all the advantage, we repeat the experiment but guess against the full dataset with the `leaked' sample $q$ removed. For example, we sample from Flirtlife without replacement and use this sample to guess Compubits, Hotmail and Rockyou normally. We then remove the users' passwords that are in the sample from the Flirtlife dataset and use the sample to guess the passwords that remain. This might also be a more accurate representation of an attackers goals: wanting to compromise additional users rather than users they already have.

\begin{figure}[ht] \hspace{-1em}
\centering
\begin{tabular}{c}
\includegraphics[width=0.55\linewidth]{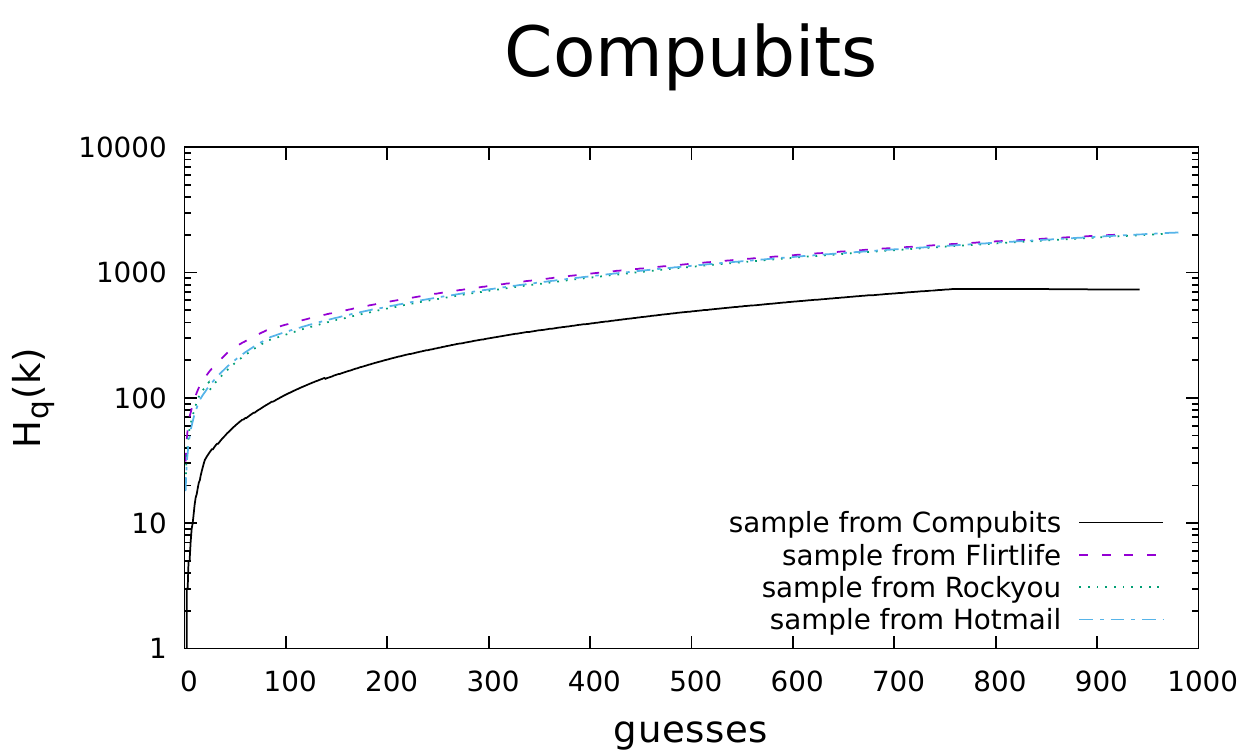}  \\
\includegraphics[width=0.55\linewidth]{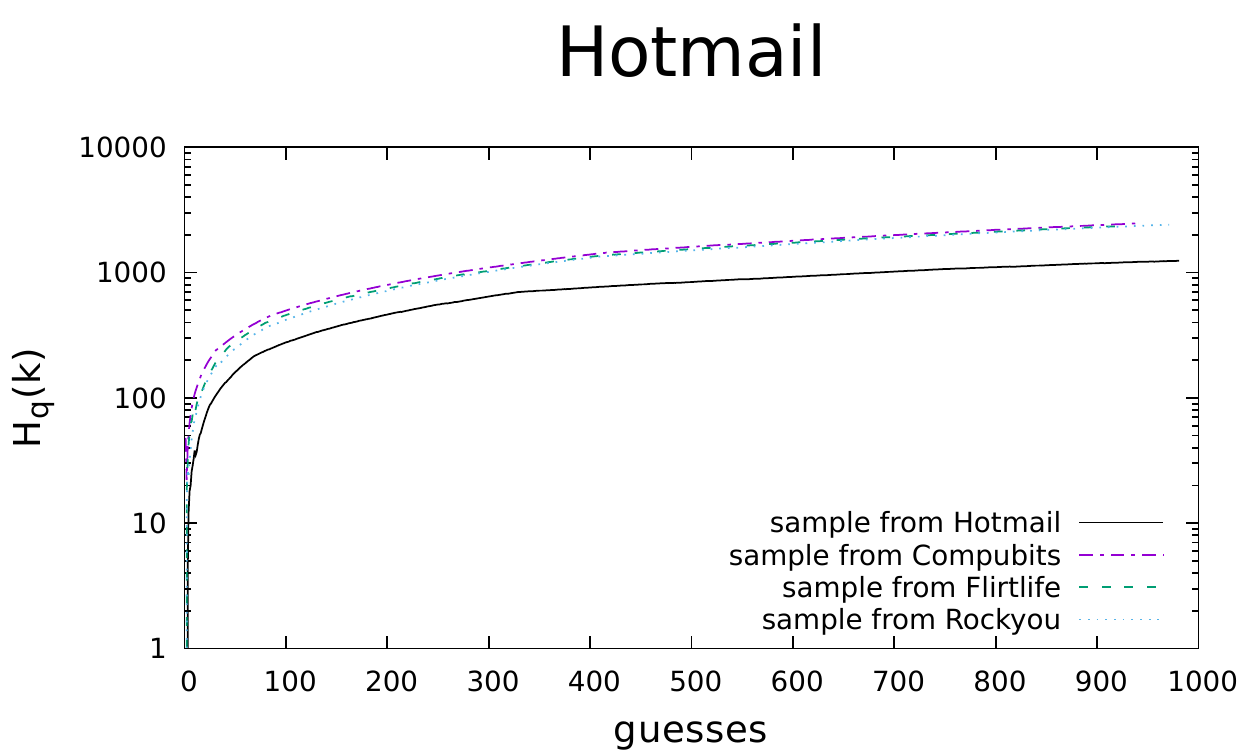}  \\
\includegraphics[width=0.55\linewidth]{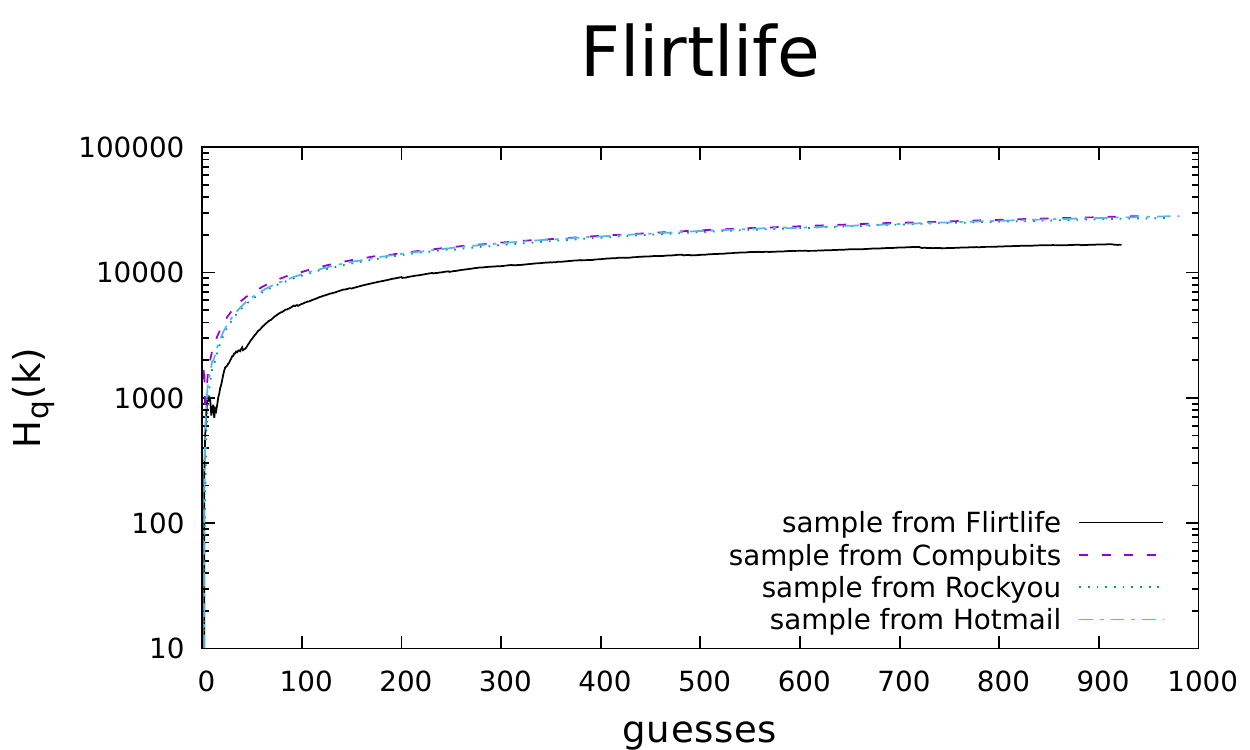}  \\
\includegraphics[width=0.55\linewidth]{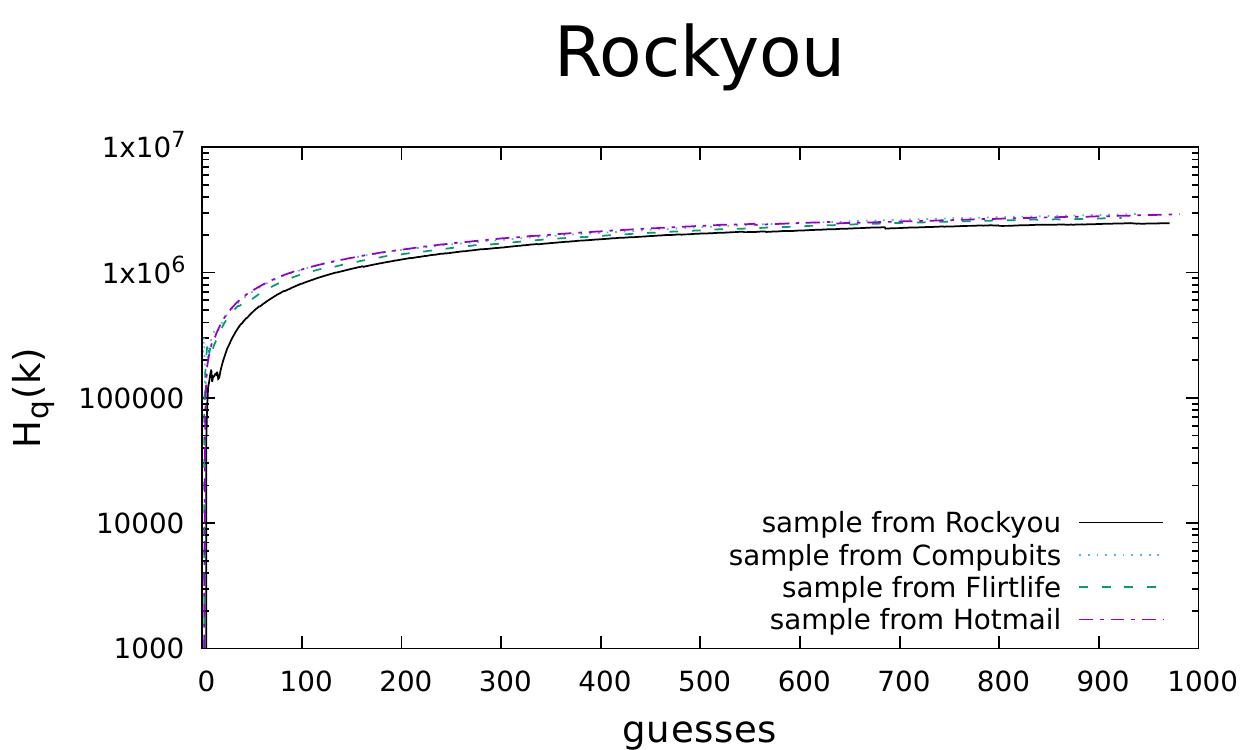}  
\end{tabular} \caption{Sampling from four different password datasets without replacement. $n=1000$ users' passwords in each sample. $\mathcal{H}_{q_t}^{1000}(g)$}  \label{fig:wo_replace}  
\end{figure}

Fig.~\ref{fig:wo_replace} compares the ability of the samples from the four different datasets at guessing each of the full password datasets. For each sample we ran multiple trials, however, for visibility in our graphs we only include the results of one trial from each sample. We will discuss the results for the other trials here. 

For compubits, hotmail and flirtlife every one of our trials demonstrated that the sample was most effective when it originally came from the dataset it is guessing. 

\begin{figure}[ht]
      \centering
      \includegraphics[width=\linewidth]{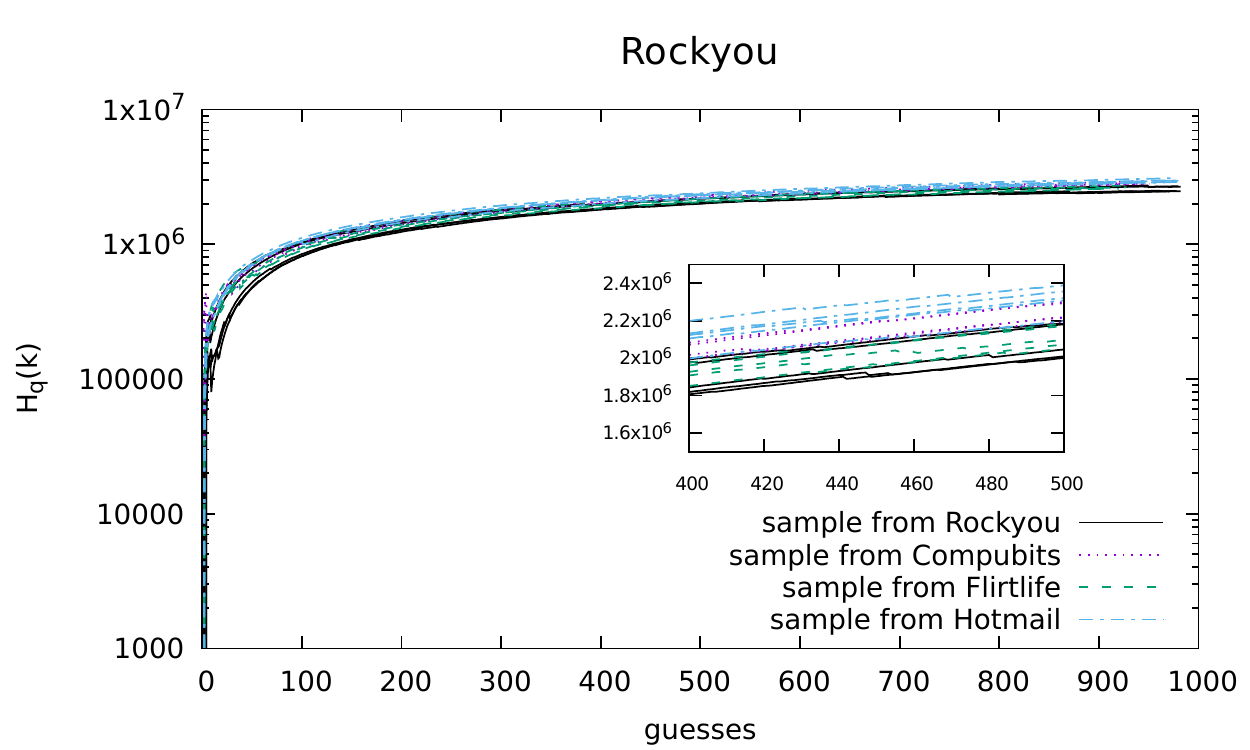}    
      \caption{Rockyou guessed using samples taken from four datasets without replacement. Trial for each sample repeated five times. Expanded x-range [$g=400$--$500$ guesses]. Expanded section does not have a log scale.}
      	\label{fig:zoom}
  \end{figure}

For the rockyou dataset the rockyou sample was not always conclusively the most effective. Fig.~\ref{fig:zoom} shows the results from 5 trials for each of the four types of samples attempting to guess the rockyou dataset. We noticed that at certain values of $g$, two of the five rockyou sample trials  were not better at guessing than some of the samples from the flirtlife dataset. To highlight this we have expanded the $400 \leq g \leq 500$ part of the plot. In this section we can see that two of the rockyou sample trials are less effective at guessing than four of the flirtlife sample trials. Therefore we conclude that for the very large rockyou dataset, the sample of leaked passwords coming from the dataset does not seem to play as important a role in the samples' ability to guess passwords.

\section{Conclusion}
In this paper we have demonstrated that what might appear to be a small leak from an organizations' password database can actually compromise a large proportion of the rest of the dataset. We believe that using our guessing function we can investigate areas such as the impact of password advice policies on guessability \cite{murray2017evaluating} \cite{kelley2012guess}.

\FloatBarrier
\bibliographystyle{ieeetr}
\bibliography{library.bib}

\begin{thebibliography}{10}

\bibitem{malone2012investigating}
D.~Malone and K.~Maher, ``Investigating the distribution of password choices,''
  in {\em Proceedings of the 21st international conference on World Wide Web},
  pp.~301--310, ACM, 2012.

\bibitem{inglesant2010true}
P.~G. Inglesant and M.~A. Sasse, ``The true cost of unusable password policies:
  password use in the wild,'' in {\em Proceedings of the SIGCHI Conference on
  Human Factors in Computing Systems}, pp.~383--392, ACM, 2010.

\bibitem{shen2016user}
C.~Shen, T.~Yu, H.~Xu, G.~Yang, and X.~Guan, ``User practice in password
  security: An empirical study of real-life passwords in the wild,'' {\em
  Computers \& Security}, vol.~61, pp.~130--141, 2016.

\bibitem{durmuth2015omen}
M.~D{\"u}rmuth, F.~Angelstorf, C.~Castelluccia, D.~Perito, and A.~Chaabane,
  ``Omen: Faster password guessing using an ordered markov enumerator,'' in
  {\em International Symposium on Engineering Secure Software and Systems},
  pp.~119--132, Springer, 2015.

\bibitem{hitaj2017passgan}
B.~Hitaj, P.~Gasti, G.~Ateniese, and F.~Perez-Cruz, ``Passgan: A deep learning
  approach for password guessing,'' {\em arXiv preprint arXiv:1709.00440},
  2017.

\bibitem{weir2009password}
M.~Weir, S.~Aggarwal, B.~De~Medeiros, and B.~Glodek, ``Password cracking using
  probabilistic context-free grammars,'' in {\em Security and Privacy, 2009
  30th IEEE Symposium on}, pp.~391--405, IEEE, 2009.

\bibitem{ethics}
D.~R. Thomas, S.~Pastrana, A.~Hutchings, R.~Clayton, and A.~R. Beresford,
  ``Ethical issues in research using datasets of illicit origin,'' in {\em
  Proceedings of the 2017 Internet Measurement Conference}, pp.~445--462, ACM,
  2017.

\bibitem{wang2016implications}
D.~Wang and P.~Wang, ``On the implications of zipf’s law in passwords,'' in
  {\em European Symposium on Research in Computer Security}, pp.~111--131,
  Springer, 2016.

\bibitem{zhang2017password}
S.~Zhang, J.~Zeng, and Z.~Zhang, ``Password guessing time based on guessing
  entropy and long-tailed password distribution in the large-scale password
  dataset,'' in {\em Anti-counterfeiting, Security, and Identification (ASID),
  2017 11th IEEE International Conference on}, pp.~6--10, IEEE, 2017.

\bibitem{fahl2013ecological}
S.~Fahl, M.~Harbach, Y.~Acar, and M.~Smith, ``On the ecological validity of a
  password study,'' in {\em Proceedings of the Ninth Symposium on Usable
  Privacy and Security}, p.~13, ACM, 2013.

\bibitem{sanov}
I.~N. Sanov, ``On the probability of large deviations of random variables,''
  tech. rep.

\bibitem{russianSanov}
\foreignlanguage{russian}{И.Н. Санов}, ``\foreignlanguage{russian}{О
  вероятности больших отклонений случайных
  величин},'' {\em
  \foreignlanguage{russian}{Математический сборник}},
  vol.~42, no.~1, pp.~11--44, 1957.

\bibitem{cover2012elements}
T.~M. Cover and J.~A. Thomas, {\em Elements of information theory}.
\newblock John Wiley \& Sons, 2012.

\bibitem{kullbackkullback}
S.~Kullback, ``Kullback--leibler divergence,''

\bibitem{murray2017evaluating}
H.~Murray and D.~Malone, ``Evaluating password advice,'' in {\em Signals and
  Systems Conference (ISSC), 2017 28th Irish}, pp.~1--6, IEEE, 2017.

\bibitem{kelley2012guess}
P.~G. Kelley, S.~Komanduri, M.~L. Mazurek, R.~Shay, T.~Vidas, L.~Bauer,
  N.~Christin, L.~F. Cranor, and J.~Lopez, ``Guess again (and again and again):
  Measuring password strength by simulating password-cracking algorithms,'' in
  {\em Security and Privacy (SP), 2012 IEEE Symposium on}, pp.~523--537, IEEE,
  2012.

\end{thebibliography}

\end{document}